\documentclass[preprint,floats,twoside,eqsecnum,aps]{revtex4}
\usepackage{graphicx}
\begin{document}

%  \markboth{E. Ferreira and F. Pereira}{Amplitudes in pp and
%  $\bar{\rm p}$p scattering}

% \catchline{}{}{}{}{}
%%%%%%%%%%%%%%%%%%%%%%%%%%%%%%%%%%%%%%%%%%%%%%%%%%%%%%%%%%%%%%%%%%%%

\title{  AMPLITUDES IN pp AND $\bar{\rm p}$p SCATTERING}
\author{ ERASMO FERREIRA }
\address{Instituto de Fisica, Universidade Federal do Rio de Janeiro,  P.O.Box 68528\\
Rio de Janeiro, RJ 22461-200, Brazil  \\
erasmo@if.ufrj.br}
\author{FL\'AVIO PEREIRA}
\address{Observat\'orio  Nacional, Rua General Jos\'e Cristino 77\\
Rio de Janeiro, RJ 20921-400, Brazil\\
flavio@on.br}
% \maketitle
%  \begin{history}
%  \received{4 May 2007}
%  \accepted{29 May 2007 }
           % \comby{(xxxxxxxxxx)}
% \end{history}
\begin{abstract}
    Solutions for the amplitudes that give accurate description
    of pp and p$\bar {\rm p}$ scattering at high energies are
    investigated, with particular attention given to the properties
    of their zeros and slopes, whose determination is required
    for the study of the Coulomb  interference region. Proper
    extrapolations of these quantities to the LHC energies are
    important for the analysis of the forthcoming experiments.
 \end{abstract}
\maketitle
  \section{Introduction}
The detailed knowledge of t-dependence of the real and imaginary
amplitudes in pp and $\bar{\rm p}$p scattering  is important for
the study of the Coulomb interference and the determination of the
collision parameters. In general, the behavior of the real
amplitude is oversimplified, with assumption that real and
imaginary slopes are identical and that the ratio $\rho$ is
independent of t. However, the present work, based on our 
analysis\cite{flavio} of the ISR, SPS and Fermilab data, shows that the
variation of the real amplitude in the Coulomb interference region
is very fast, with a slope about twice that of the imaginary part.
The  forthcoming  experiments at RHIC and LHC,\cite{totem}
respectively at 200-500 GeV and 14 TeV,  make urgent the
investigation of   the structure of the amplitudes in the low
$|t|$ region.\cite{bourrely}

The real forward slope is obviously  connected with the value of t at
which this amplitude passes through zero. We find this connection,
fulfilling the expectation from the theorem by A. Martin\cite{martin}
that the real part has a zero that approaches the forward direction
as the energy increases.

Our work covers the full t-range, obtaining
also the zeros of imaginary amplitudes and the shapes
of the dips in the cross sections of pp and
 $\bar{\rm p}$p scattering.
\section{Amplitudes, Slopes and Zeros}
We have constructed an accurate description of the t-dependence
of the amplitudes, with smooth energy dependence in the parameters,
with particular attention to the determination of slopes and zeros.

\begin{figure}[th] 
\includegraphics*[ width=6.5cm] {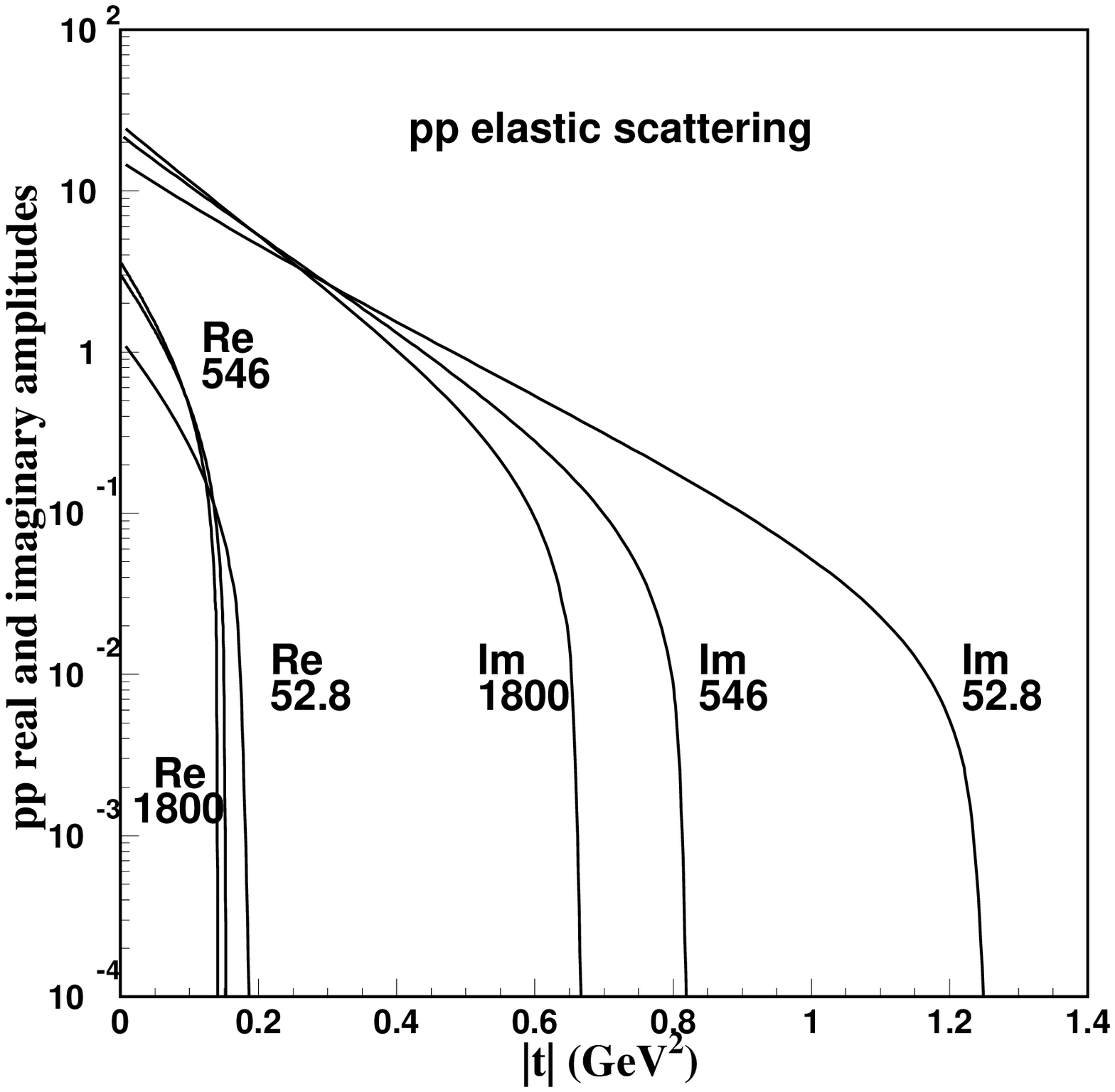} \hfill
\includegraphics*[ width=6.5cm] {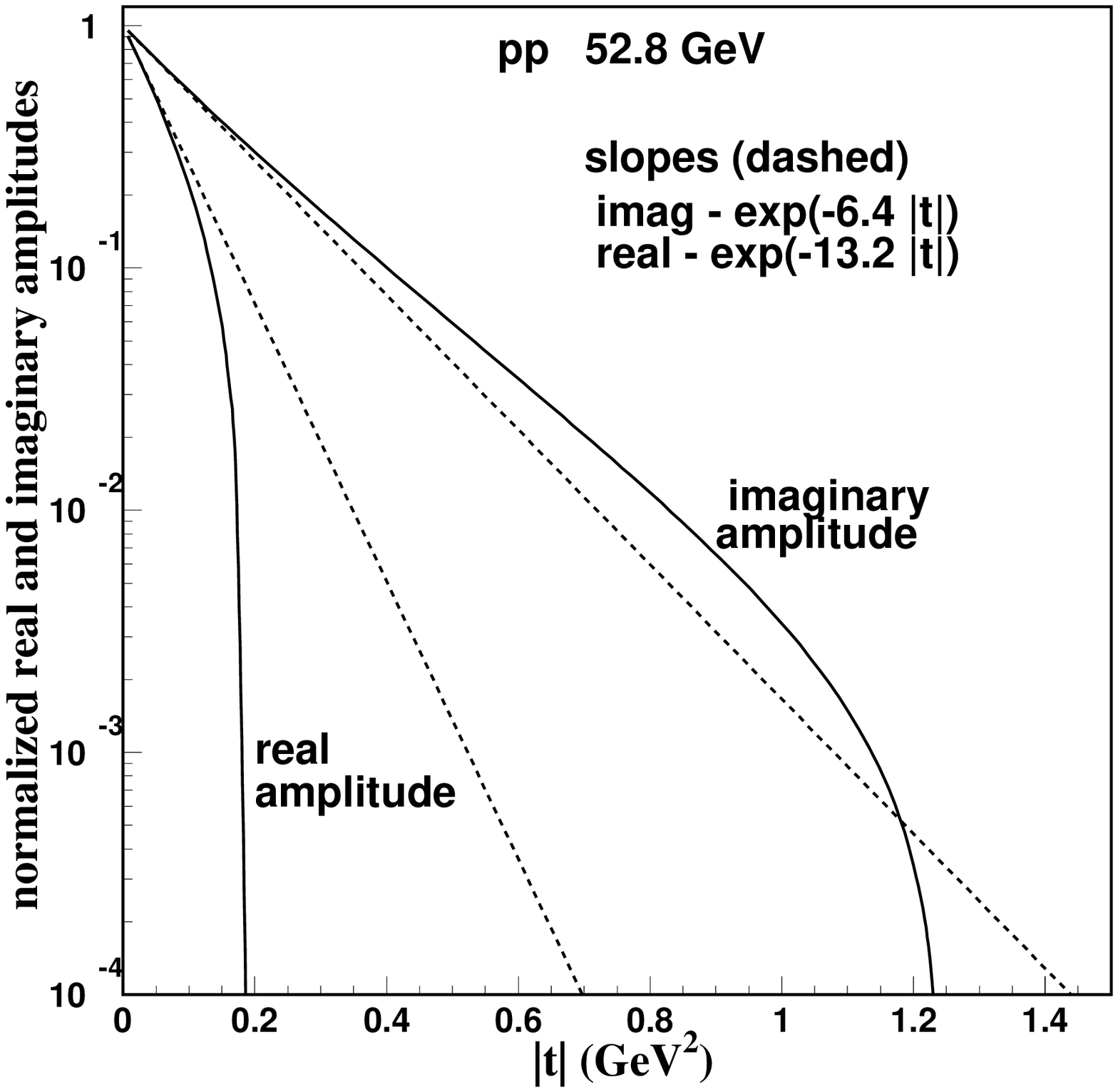}
\caption{
  a-Left: Real and imaginary amplitudes in forward directions
  for pp scattering at 52.8 GeV, and $\bar{\rm p}$p  at 541 and 1800  GeV,
  obtained through  analysis of the data on differential cross sections.
  b-Right: Detail of the pp amplitudes at 52.8 GeV, normalized to one at
  t=0, with determination of the exponential slopes. The imaginary
  amplitudes show upwards curvature before they start decreasing to zero.}
\label{ampl}
\end{figure}

\begin{figure}[ht]
\includegraphics*[ width=6.5cm] {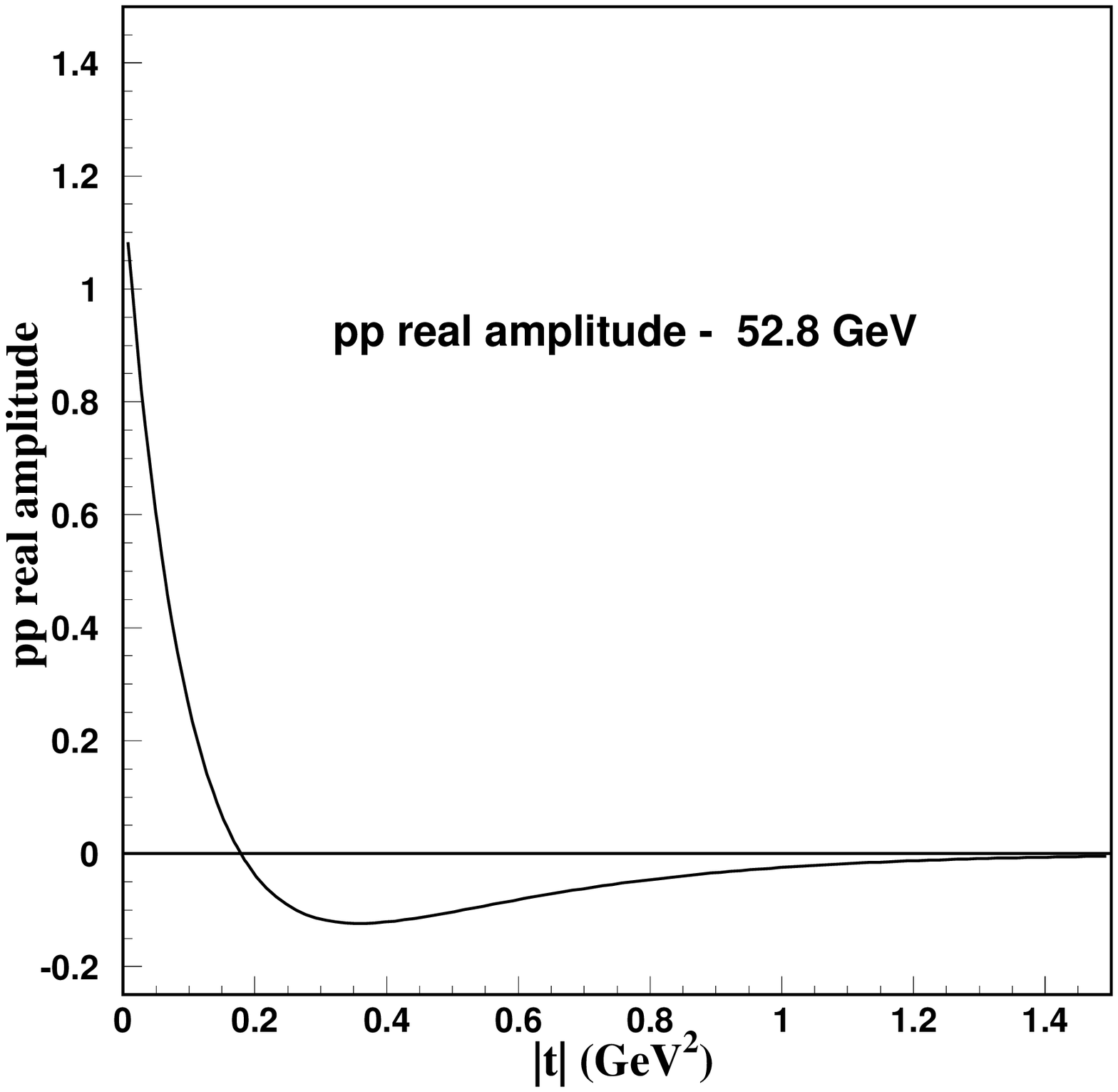} \hfill
\includegraphics*[ width=6.5cm] {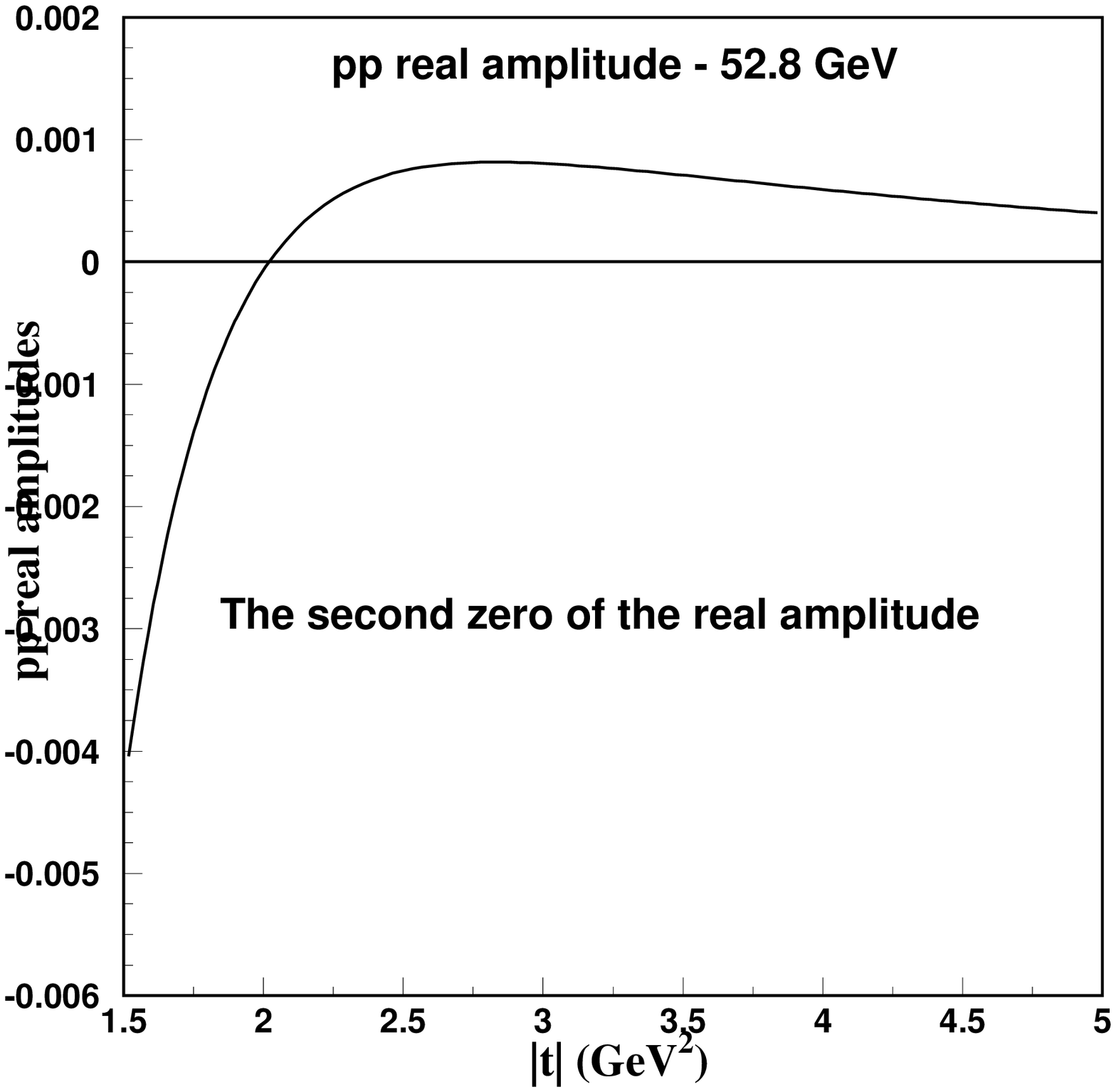}
\caption{
 Real amplitude of pp  scattering at 52.8 GeV.
  The two separate  figures point to different parts of the
$|t|$ range where the first and second zeros are located.}
\label{real52fig}
\end{figure}

 The amplitudes have peculiar shapes as functions
of $|t|$, with approximate exponential behaviour only for very small
$|t|$, and then turning fast towards zero.
The results for some energies (52.8 GeV in pp, and 541 and 1800 GeV
for $\bar {\rm p}$p ) are shown in Fig. \ref{ampl}.
    Theoretical arguments\cite{martin} about the existence of a real
zero close to $ |t|=0 $  at
high energies are confirmed by our construction. The real pp amplitude
 has a second zero, while in $\bar{\rm p}$p only the
first zero at small $|t|$ is observed.\cite{flavio} The pp
case is drawn in  Fig. (\ref{real52fig}), where the real amplitude at
$\sqrt{s}=52.8$ GeV data is shown for small and for large $ |t| $,
so that we can see where the second zero in the pp channel occurs.
This second zero, being close to the zero of the imaginary amplitude,
stresses the dips observed in the pp differential cross sections,
but not those of the $\bar{\rm p}$p channel,\cite{flavio} due to
a difference in the sign of the real tail.
% The elastic differential cross section is given by
% \begin{equation}
% \label{diff}
% \frac{d\sigma^{el}}{d|t|}=\frac{1}{16\pi s^2}|F(s,t)|^2
% =\frac{1}{16\pi s^2}\bigg[ [{\rm Re}F(s,t)]^2+[{\rm Im}F(s,t)]^2 \bigg]
% \end{equation}

With real and imaginary amplitudes  described at low $|t|$
by exponential slopes, we write   for each channel,
$ {\rm Re} F(s,t)={\rm Re} F(s,0) \exp (-{B^R} |t|/2)$,
$ {\rm Im} F(s,t)={\rm Im} F(s,0) \exp (-{B^I} |t|/2) ~  $
and the experimental slope $B$ is given by
$ B(s)=\big(\rho^2(s) B^R+B^I\big)/\big(\rho^2(s)+1\big)~ . $
We determine separately  $B^R$ and $B^I$  for each
channel, and observe the approximate relation
$B^R \approx 2 B^I$  above 30 GeV.

The exponential slopes at $|t|=0$   for  52.8 GeV are shown in
% Fig. \ref{slopefig}
Fig. \ref{ampl}, right hand side, where we plot the pp amplitudes
 normalized to one  at $|t|=0$ . We
observe that the concept of exponential slope is limited to a small
$|t|$ range, with strong characteristic curvatures of the real and
imaginary amplitudes appearing very soon in the $|t|$ scale. It is
remarkable that the imaginary amplitude has a positive (going upwards)
curvature at the beginning, before it starts running  fast to zero.

\begin{figure}[ht]
\includegraphics*[ width=6.5cm]{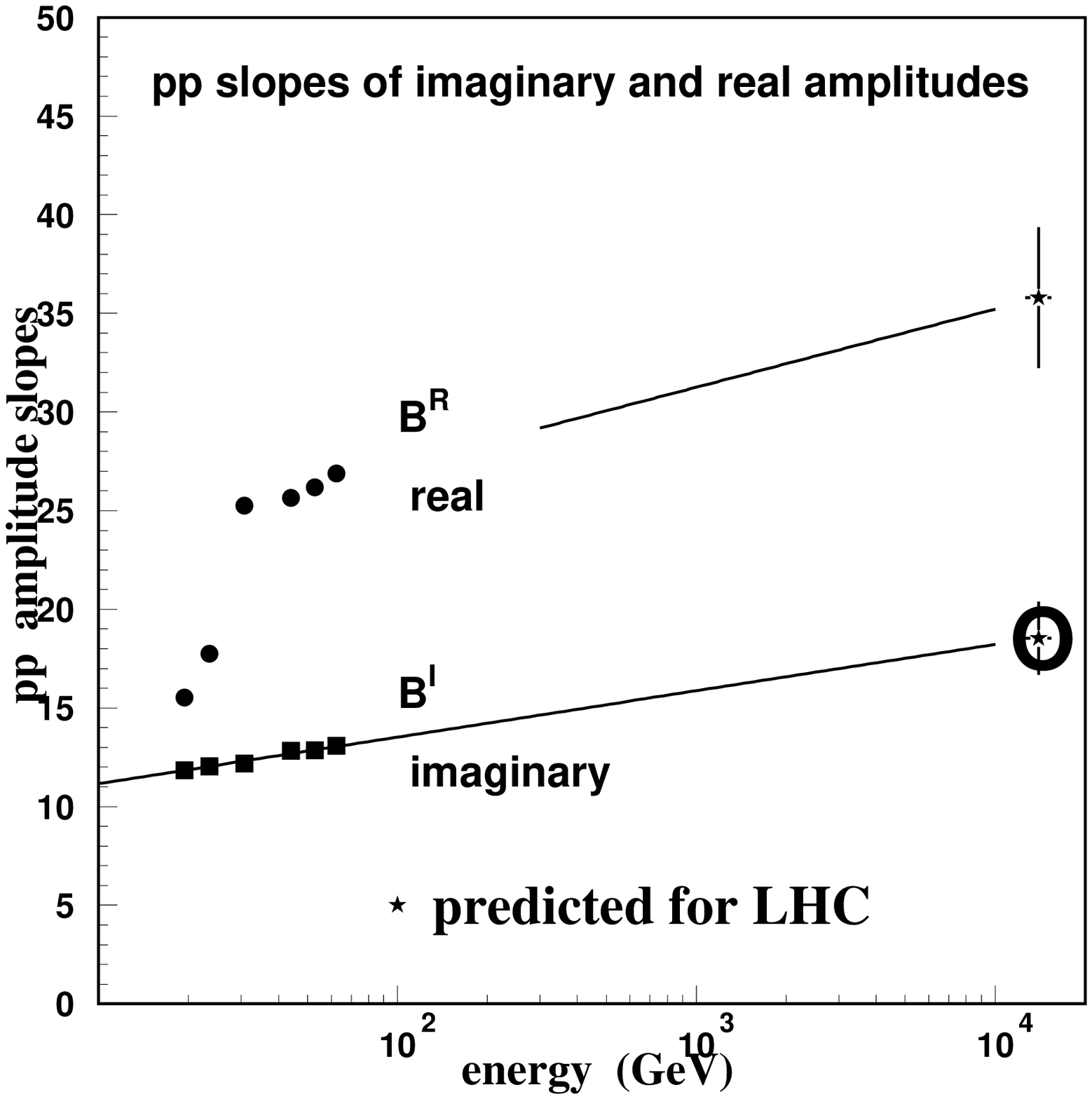} \hfill
\includegraphics*[ width=6.5cm]{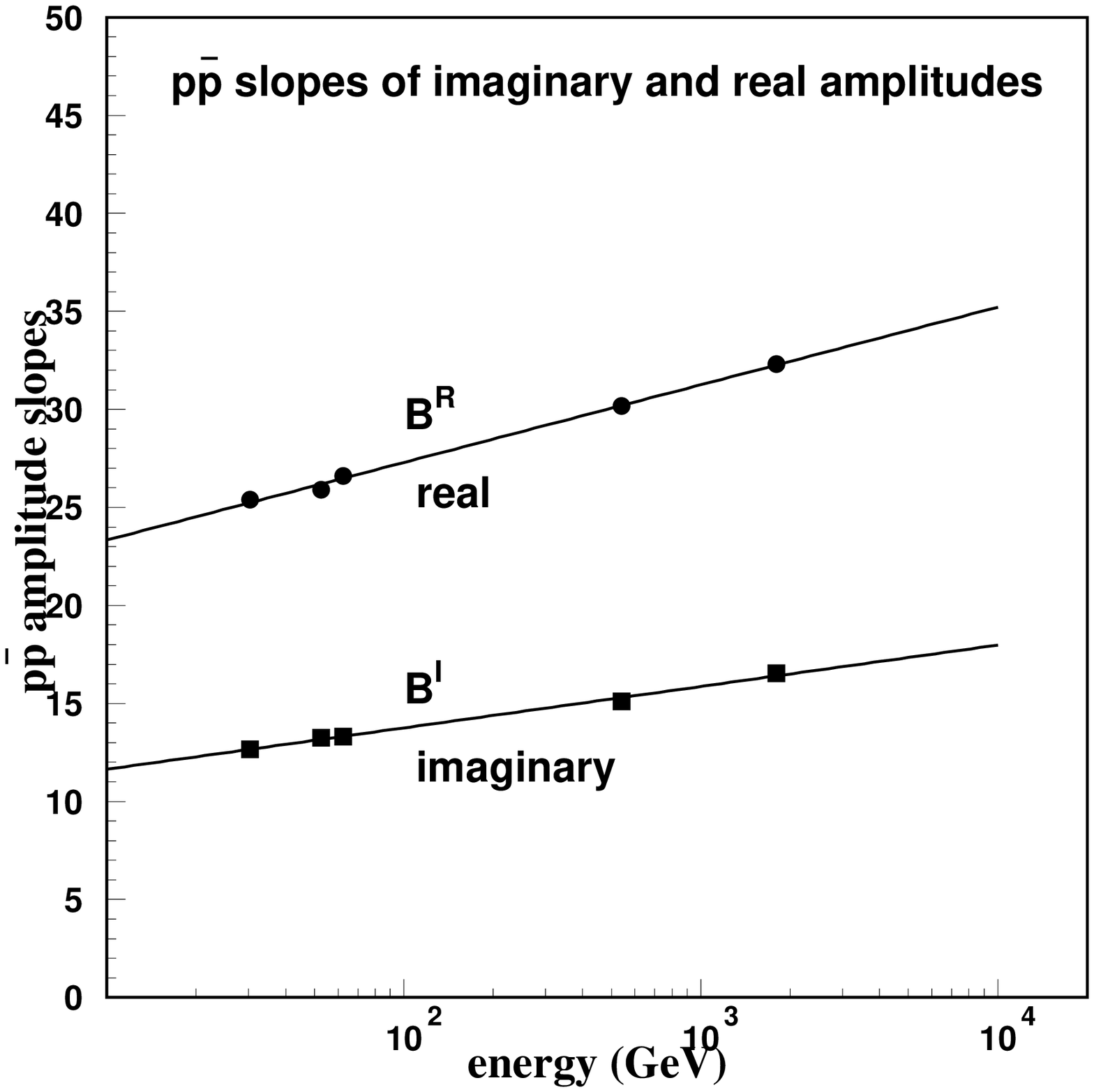}
\caption{
 Slopes of imaginary and real amplitudes in pp and ${\bar{\rm p}{\rm p}} $
 scattering, extracted from our parametrizations of  the amplitudes. }
\label{br_bi}
\end{figure}

  These results for the slopes  are shown in Fig. \ref{br_bi}.
For $B^I$ we have a simple parametrization
$ B^I_{\rm pp}=B^I_{\bar{\rm p}{\rm p}}=B^I(s)=8.6255+1.0372 ~  \log{\sqrt{s}} $ in ${\rm GeV}^{-2}$,
valid for both pp and $\bar{\rm p}$p channels, with a  prediction
 $B^I=18.53 ~  {\rm GeV}^{-2}$ for the LHC energy  $\sqrt{s}=$ 14 GeV.
 The description is not so simple for the real  slope. Here
the pp and  ${\bar{\rm p}{\rm p}} $ cases are different at the
lower energies, and an extrapolation to higher energies is not neat
for pp. For ${\bar{\rm p}{\rm p}} $ , using the 541 and 1800 GeV
points,  we guess a linear $\log s$ behavior
$ B^R_{\bar{\rm p}{\rm p}}=19.4314+1.7569 ~ \log{\sqrt{s}}$
in ${\rm GeV}^{-2}$.

The role of $ B^R$ in the determination of the collision parameters
can be seen in the work of Gauron, Nicolescu and 
Selyugin,\cite{nicolescu} who  find a point
$t_{\rm min}$ where the real amplitude is compared to the Coulomb part.
They study pp scattering at 52.8 GeV, suggesting a
lower value for $\rho$ than 0.077 determined experimentally. However,
to write the real amplitude at the matching point $t_{\rm min}$ ,
they use the exponential factor $ \exp{-B|t_{\rm min}|/2}$ with the
experimental slope $B=12.87 ~ {\rm GeV}^{-2}$. However, according to our
work, they should use the faster  slope $ B^R=26 ~ {\rm GeV}^{-2}$
of the real amplitude. Then the new value that they suggest for
$\rho$ is not properly evaluated.
 \section{Theorem on the first real zeros}
\begin{figure}[ht]
\includegraphics*[ width=6.5cm]{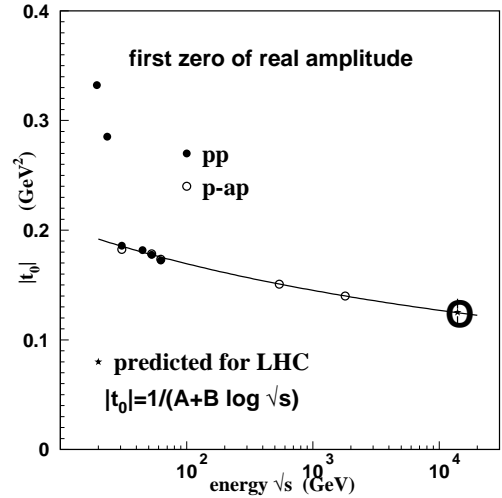} \hfill
\caption{ Values $|t_0|$ of the zeros of the real amplitude in pp
and $\bar{\rm p}$p scattering.}
\label{zeros} \end{figure}

A theorem by A. Martin\cite{martin} predicts that in pp and
$\bar{\rm p}$p scattering  the real part has a change of sign at a
point $|t_0|$ which moves to the forward direction as the energy
increases. The paper  suggests real zeros of the form
$ |t_0|= 1/(A+B \log s) ~ , $
as we find with our description of amplitudes, as shown in
Fig. \ref{zeros}.
%  \section*{Acknowledgements}

{\bf Acknowledgements. }  The authors are grateful to CNPq (Brazil) for 
support of their scientific research.
 

\begin{thebibliography}{0}
\bibitem{flavio} F. Pereira and E. Ferreira,  {\it Phys. Rev.} {\bf D59} (1999) 014008
  ; {\it Phys. Rev.} {\bf D61} (2000) 077507.
\bibitem{totem} G. Anelli et al., ``TOTEM physics'', hep-ex/0602025.
% \bibitem{cudell} J.R. Cudell et al., Phys. Rev. {\bf D65} (2002), 074024
%       and Phys. Rev. Lett. {\bf 89}, 201801 (2002).
\bibitem{bourrely} C. Bourrely et al., ``Why the real part of the 
proton-proton forward
scattering amplitude should be measured at the LHC'', hep-ex/0511135.
\bibitem{martin} A. Martin,  {\it Phys. Lett.} {\bf  B404} (1997) 137.
\bibitem{nicolescu} P. Gauron, B. Nicolescu, O.V. Selyugin,
     {\it Phys. Lett.} {\bf B629} (2005) 83.
\end{thebibliography}
\end{document}